\documentclass[conference]{IEEEtran}
\IEEEoverridecommandlockouts
\usepackage{cite}
\usepackage{amsmath,amssymb,amsfonts}
\usepackage[algo2e]{algorithm2e}
\usepackage{algorithm,algpseudocode}
\usepackage{graphicx}
\usepackage{textcomp}
\usepackage{xcolor}
\usepackage{xcolor}
\def\BibTeX{{\rm B\kern-.05em{\sc i\kern-.025em b}\kern-.08em
    T\kern-.1667em\lower.7ex\hbox{E}\kern-.125emX}}
\begin{document}

\title{MEC-Intelligent Agent Support for Low-Latency
Data Plane in Private NextG Core}

\author{\IEEEauthorblockN{Shalini Choudhury}
\IEEEauthorblockA{\textit{}
\textit{WINLAB, Rutgers University}\\
shalini@winlab.rutgers.edu}
\and
\IEEEauthorblockN{Sushovan Das}
\IEEEauthorblockA{\textit{}
\textit{Rice University}\\
sd68@rice.edu}
\and
\IEEEauthorblockN{Sanjoy Paul}
\IEEEauthorblockA{\textit{}
\textit{Accenture Labs}\\
sanjoy.paul@accenture.com}
\and
\IEEEauthorblockN{Prasanthi Maddala, \\ Ivan Seskar, \\ Dipankar Raychaudhuri}
\IEEEauthorblockA{\textit{}
\textit{WINLAB, Rutgers University}\\
prasanti,seskar,ray@winlab.rutgers.edu}
}

\maketitle

\begin{abstract}
Private 5G networks will soon be ubiquitous across the future-generation smart wireless access infrastructures hosting a wide range of performance-critical applications. A high-performing User Plane Function (UPF) in the data plane is critical to achieving such stringent performance goals, as it governs fast packet processing and supports several key control-plane operations. Based on a private 5G prototype implementation and analysis, it is imperative to perform dynamic resource management and orchestration at the UPF. This paper leverages Mobile Edge Cloud-Intelligent Agent (MEC-IA), a logically centralized entity that proactively distributes resources at UPF for various service types, significantly reducing the tail latency experienced by the user requests while maximizing resource utilization. Extending the MEC-IA functionality to MEC layers further incurs data plane latency reduction. Based on our extensive simulations, under skewed uRLLC traffic arrival, the MEC-IA assisted \textit{bestfit UPF-MEC} scheme reduces the worst-case latency of UE requests by up to $77.8\%$ w.r.t. baseline. Additionally, the system can increase uRLLC connectivity gain by $2.40\times$ while obtaining $40\%$ CapEx savings.
\end{abstract}


\begin{IEEEkeywords}
Data plane, low latency, cloud-native core, private 5G network, mobile edge computing.
\end{IEEEkeywords}
\vspace{-4mm} 
\section{Introduction}

\begin{figure}[t]
\centering
\includegraphics[width=0.95\columnwidth]{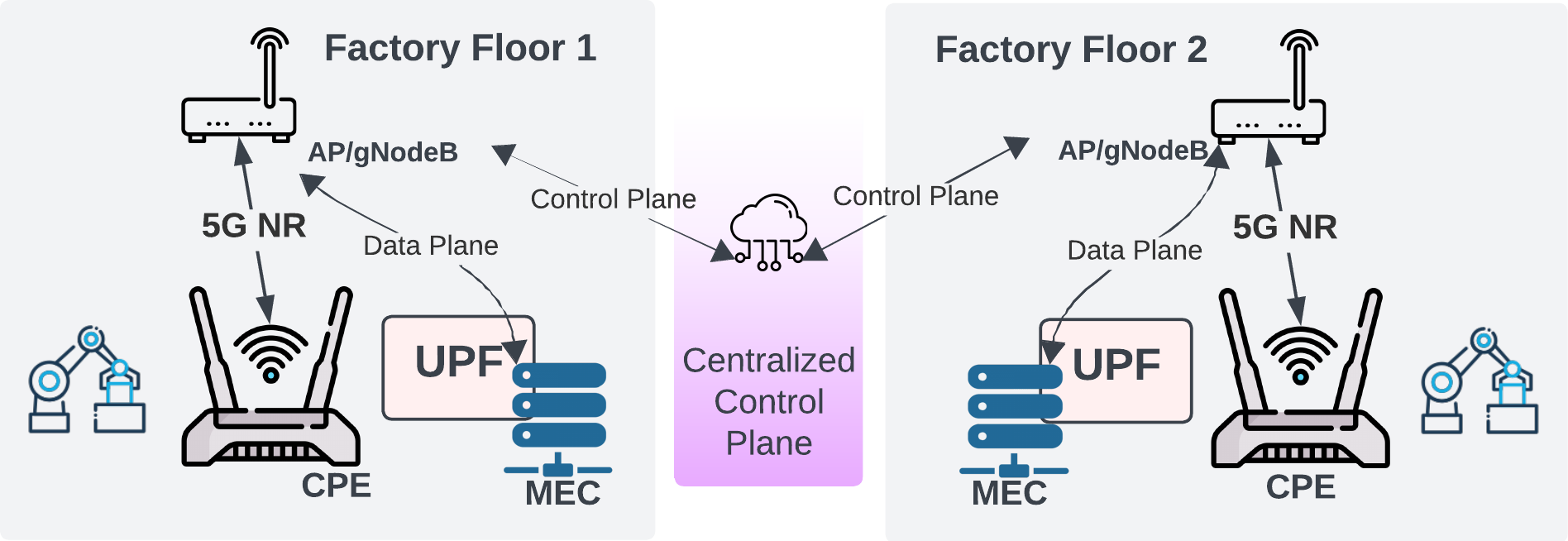}
\caption{Standalone Private 5G Implementation in Smart Factory Ecosystem.}
\vspace{-7mm}
\label{fig:Intro-P-5G}
\end{figure}

5G networks offer modularity, programmability, and flexibility to cater to a diverse range of application requirements. The unique demands of various industries, including customized performance and privacy needs, have catalyzed the development of private 5G networks. 
Among various architectures, the private 5G \textit{Stand-Alone} (SA) Deployment model allows for an independent system, free from dependencies on legacy networks, offering convenient technical enhancements revolving around softwarization, cloudification, and increased modularity. This model allows operators to have more flexibility in working with third-party service providers, which potentially improves the quality of service (QoS) of performance-critical applications. This work focuses on SA private 5G in the realms of low-latency applications such as smart manufacturing, where the user data plane terminates locally with pertinent network functions and services located in the nearby Mobile Edge Cloud (MEC). As real-world industrial manufacturing may be distributed over multiple facilities and large geographical distances, an approach for interconnecting sites is shown in Figure \ref{fig:Intro-P-5G}. The network's control plane resides at a central site, while the data plane is distributed across all the factory floors. The rationale behind this design choice is the ease of centralized management of network-wide mobility, authentication, and policies while the distributed data plane can support processing the data traffic closer to its source or destination.

Smart factory ecosystem embodies a wide range of services that differ significantly in terms of requirement including a high throughput, ultra-low processing latencies, and stringent QoS enforcement. The mobile packet core, which connects the radio access network to external networks, comprises of 1) a control plane that processes signaling messages and 2) a data plane that forwards user traffic. The UPF is the major entity in the data plane and has a significant impact on the performance that users would perceive with 5G. 
Based on the standalone private 5G prototype on the COSMOS testbed \cite{COSMOS}, we observe that the compute utilization of UPF escalates abruptly as the number of end-user connectivity scales up. Such data-plane compute bottleneck at UPF would result in significant QoS degradation (such as high tail latency) of the diverse performance-critical applications. The situation can become worse if the traffic arrival across different UPFs is highly skewed, which would create a ripple effect of network and compute resource congestion at the MEC layer.

In this paper, we leverage MEC-Intelligent Agent (MEC-IA), a logically centralized entity hosted in the MEC platform that can potentially remove the bottleneck and realize application-aware fine-grained resource assignment in the data plane. MEC-IA periodically monitors UPFs’ and MECs’ utilization across all the critical services and thus realizes proactive and dynamic resource provisioning in real-time. Previous research has primarily focused on multipath load balancing in the data plane to fully utilize the available bandwidth for group of flows \cite{HULA, HULA2}. However, the MEC-IA approach attempts to perform an efficient, proactive assignment of UE to a UPF 
for setting up the PDU session, 
which precedes the commencement of data flow. Hence, this work is precursory and complementary to software-defined data plane management. The contributions are summarized as follows:
\begin{enumerate}
    \item Implement standalone end-to-end deployment of a private 5G network to benchmark compute and network resource utilization by increasing the emulated UEs.
    \item Dynamically and proactively allocate resources to maximize overall system utilization and minimize data plane delay across all the services and the UPFs. 
    \item Extend the data plane resource management and orchestration to the integrated local MEC in the private 5G system to avoid a bottleneck in the MEC layer.
    \item Leverage MEC-Intelligent Agent (MEC-IA), hosted in a MEC platform in the smart factory space that orchestrates and manages resource distribution precisely and rapidly to provide a well-classified QoS for real-time communications while also performing slice provisioning at each UPF as per UE’s service subscription.

\end{enumerate}

\vspace{-2mm}
\section{Motivation}

\vspace{-3mm}
\begin{figure}[h]
\centering
\includegraphics[width=0.95\columnwidth]{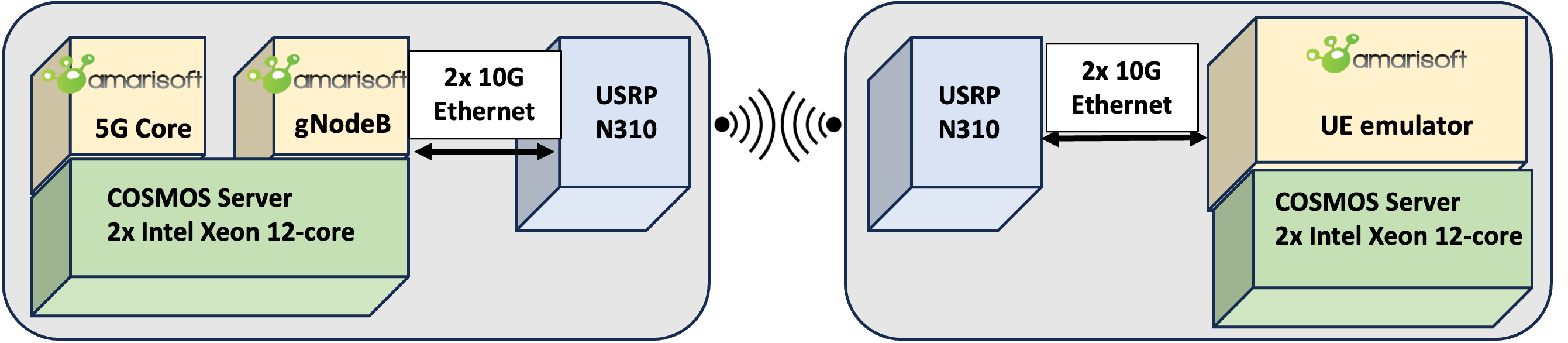}
\caption{Prototype Private 5G set-up in COSMOS testbed.}
\vspace{-5mm}
\label{fig:amarisoft-my}
\end{figure}

\subsection{Private 5G Implementation: Data Plane Bottleneck at UPF}
\label{proto}

To gain insight into the performance of 3GPP standardized private 5G infrastructure similar to smart factory network settings, we implemented a standalone private 5G prototype on the COSMOS testbed \cite{COSMOS} using Amarisoft's software \cite{amarisoft}. As seen in Figure \ref{fig:amarisoft-my}, the 3GPP-complaint gNodeB is supported by Software Defined Radio (SDR) offered by the testbed and is assigned dedicated processor cores for its operation. 
\begin{figure}[t]
\centering
\includegraphics[width=0.98\columnwidth]{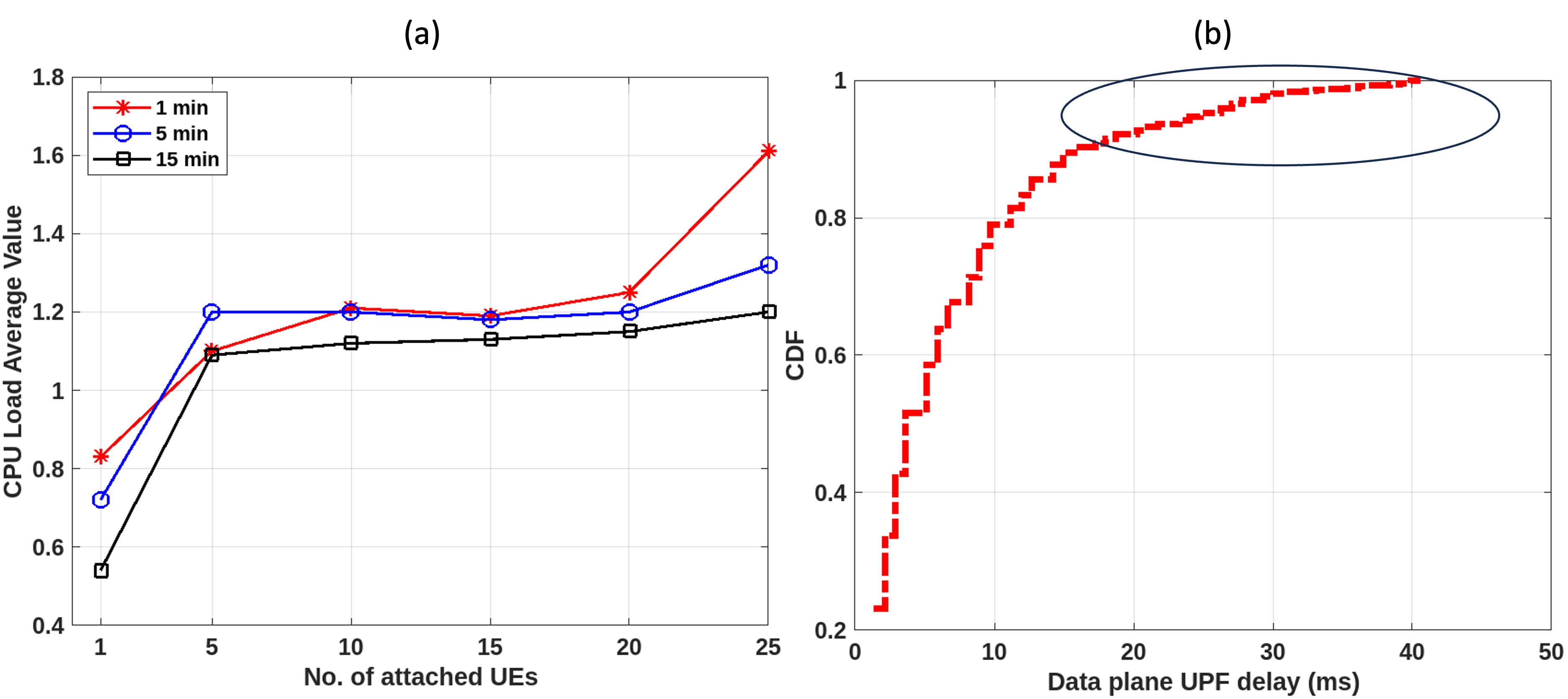}
\vspace{-3mm}
\caption{Load-average for UPF utilization vs. number of attached UEs.}
 \vspace{-3mm}
\label{fig:amarisoft}
\end{figure}
\textbf{Observation:} As the number of emulated UEs are scaled up in parallel, we observe that the bandwidth per connection drops as a function of the increase in the number of UE connections. Figure \ref{fig:amarisoft}(a) shows the load-average of the processing core that hosts the containerized UPF. The load-average can be considered a measure of both CPU utilization and waiting tasks. With one UE connected, the 1, 5, and 15-minute load averages are 0.83, 0.72, and 0.53, respectively. Implying that for one minute, the UPF reports 83\% CPU utilization. Similarly, for 5 and 15 minutes, the UPF utilization is 72\% and 53\%, respectively. With five UEs connected to the network, a 1-minute load-average of 1.10 indicates the CPU was fully used with an extra 10\% of UPF processes waiting. At 25 UEs, 61\% of UPF processes awaited CPU time. Therefore with the assigned UPF resources, it struggles with processing control and data plane messages for more than two simultaneous UE connections. CPU usage peaks between 1-5 and 20-25 UEs, causing delays in processing UE connections at the UPF.

\subsection{Challenges in Supporting Smart Factory Use Cases}
Smart factory use cases necessitate high data rates and low latency. Autonomous robots communicate periodically for motion and machine control, while sensors require high connection density for periodic measurements. These use cases can be grouped into Ultra-reliable low-latency communication (uRLLC), Massive machine-type communications (mMTC), and Enhanced mobile broadband (eMBB).

In the integrated framework of autonomous robots and sensors, rising mobile-traffic intensifies computation and offloads traffic to edge cloud \cite{usecase1}. A simulation was designed for large-scale Private 5G data plane analysis with multiple UPFs and different QoS traffic types, which is currently a limitation in the Amarisoft prototype setup. In Figure \ref{fig:amarisoft}(b), traffic of different QoS arrives at the base station, and the SMF chooses a UPF based on UE location, capability, and load. The median latency is around 5 ms, meeting ultra-low latency needs. However, the $95^{th}$ percentile latency reaches 30 msec, $6\times$ the median. This tail latency spike captured in simulation is the effect of CPU load spikes (Figure \ref{fig:amarisoft}(a)). Hence it is imperative to efficiently analyze the type of UE request and perform dynamic resource allocation to achieve low latency communication, high bandwidth and reliability.

\begin{figure}[t]
\centering
\includegraphics[width=0.7\columnwidth]{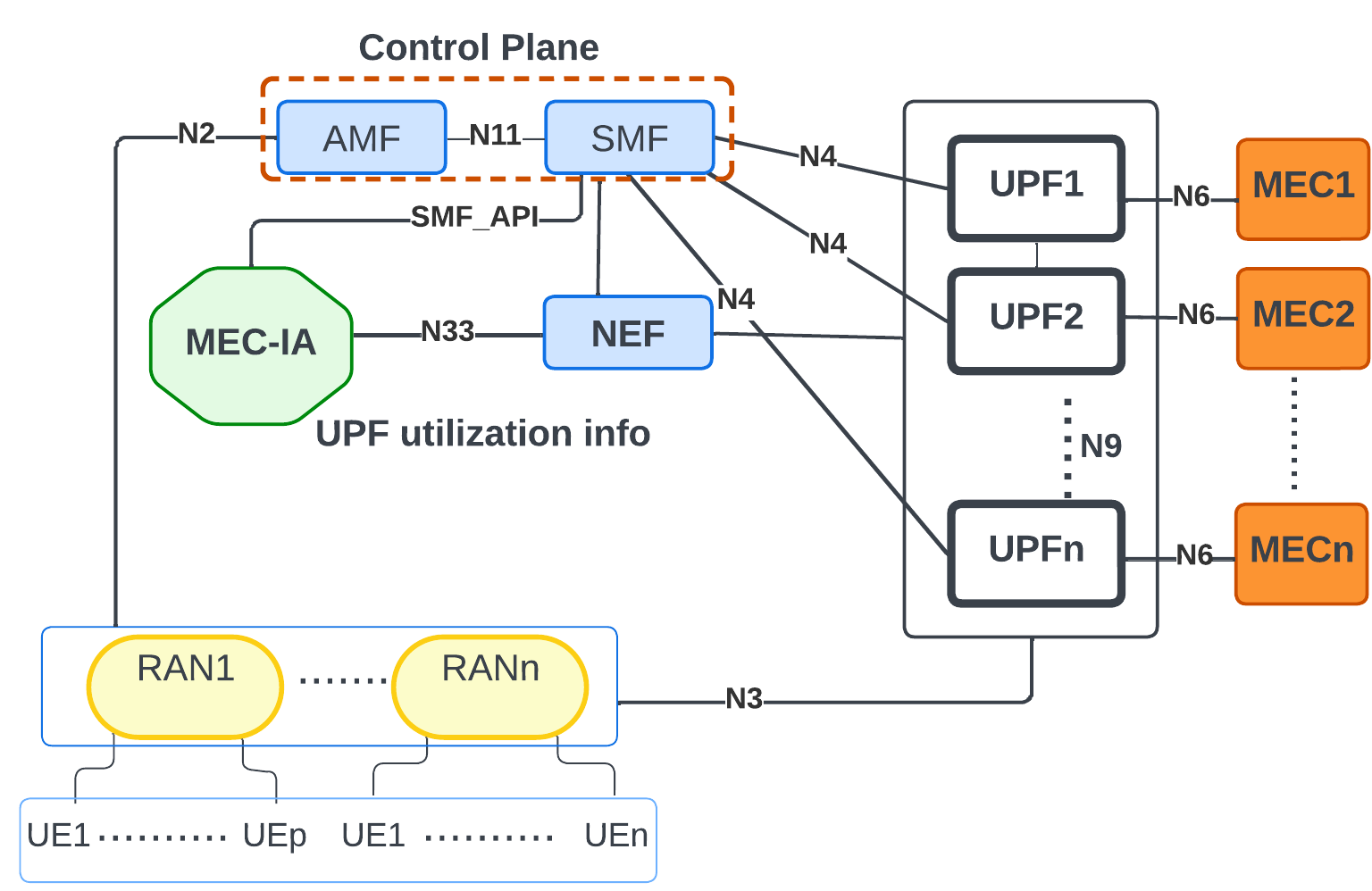}
\caption{MEC-IA assisted private 5G network architecture.}
\vspace{-6mm}
\label{fig:MEC-IA}
\end{figure}

\vspace{-2mm}
\section{MEC-IA Assisted Private 5G Architecture}


In private 5G settings, UPFs are placed in close proximity to the RAN and are geographically co-located with the MEC deployments to address the limitations of a backhaul bandwidth bottleneck.
UPFs handle all user-plane traffic for egress into data networks like the public internet or local MEC. As depicted in Figure \ref{fig:MEC-IA}, a UE connects to the UPF through RAN using the GTP Protocol \cite{N3-N9}. UPFs have multiple roles, including PDU session control, packet routing, traffic reporting, uplink verification, and transport packet marking. They are pivotal in managing control messages and data plane traffic and cater to various network slices like uRLLC, eMBB, and mMTC to meet diverse QoS demands.

This highlights the importance of implementing redundancy and efficient fault management mechanism in network design to reduce the bottleneck at the UPF. Therefore, to ensure service continuity and system reliability, this paper leverages a logically centralized \textit{\textbf{M}obile \textbf{E}dge \textbf{C}ompute-\textbf{I}ntelligent \textbf{A}gent} (MEC-IA) that monitors information collected from network exposure function (NEF) over a 3GPP standardized N33 interface \cite{N3-N9}. The information includes UPFs' utilization across all the critical services and compute and network utilization for all the MECs in the private 5G network. MEC-IA executes Algorithms \ref{alg:best_fit} (Section \ref{Algo}) to select a \textit{Bestfit UPF-MEC} pair to minimize global data plane latency by maximizing resource usage. 
The pairing of \textit{bestfit UPF} with \textit{bestfit MEC} can be easily realized in the standalone private 5G scenario since the MEC platform is integrated with the network such that it is co-located with the UPF, which makes rerouting between UPF-MEC pairs potentially simpler.  

\vspace{-4mm}
\section{System Model}
\label{Algo}

\vspace{-1mm}

The system considers a two-tier computation model for the data-plane evaluation, Figure \ref{fig:sim-design}. In the first layer, heterogeneous compute \textit{UPFs = \{1,..., U\}} caters to uRLLC, eMBB, mMTC and regular traffic. In the second layer, heterogeneous \textit{MECs = \{1,..., M\}} are co-located with the UPF. Each UPF  has limited resources to support different service categories, as seen in Figure \ref{fig:sim-design}. Similar is the case with MEC, except that regular traffic is sent to the data network when it egresses from the UPF, while all the other categories of services (uRLLC, eMBB, and mMTC) are processed without any biases. 

\subsubsection{Service Based Queue}

Each slice has different QoS requirements in terms of bandwidth, latency (jitter), packet loss rate and reliability. Thus, the system model adopted in this work considers compute queue corresponding to the QoS requirements. For example, Figure \ref{fig:sim-design} (a) and (b) shows four compute queues at all UPFs for uRLLC, eMBB, mMTC and regular traffic. The queue length at the UPF for each service is determined by the computing capabilities allocated for that service at that UPF. The queue length of the uRLLC, eMBB, mMTC, and regular traffic is proportional to the ratio of the average arrival rate to the average service rate of a particular QoS. Per standards, network slicing ends at the UPF, and at the MEC, there are no distinct queues for each QoS type, adhering to first come first serve queueing discipline.

\vspace{-4mm}
\subsection{Performance Metrics}
\vspace{-1mm}
\subsubsection{Compute Delay at UPF}

We model the compute delay (Eqn. \ref{eq:1} and Eqn. \ref{eq:2}) of a UE request for a given QoS type ($QoS = 1,2,3,4$ meaning uRLLC, eMBB, mMTC and regular traffic respectively)  served at $UPF[i]$ ($D_{UPF}[i][QoS]$)  as a function of service queue length ($Q_{UPF}[i][QoS]$), compute capacity of UPF reserved for that service ($C_{UPF}[i][QoS]$), and compute server utilization ($S_{UPF}[i][QoS]$). The 
compute capacity of the UPF is modeled as the function of its CPU processing speed i.e., Execution Time Per Bit ($ETPB_{UPF}[i])$, bytes processed in UPF per UE ($Bytes_{UPF}[i]$) and the resource factor ($\alpha[i][QoS]$) dictating the fraction of compute resource reserved for that given service on the UPF (Eqn. \ref{eq:3}). 
\begin{equation}
    \small
    \setlength{\belowdisplayskip}{0pt}
    \setlength{\abovedisplayskip}{0pt}
    D_{UPF}[i][QoS] = (\frac{(Q_{UPF}[i][QoS] + 1 - S^{\Delta}_{UPF}[i][QoS])}{C_{UPF}[i][QoS]}+1)*\delta
    \label{eq:1}
    \end{equation}

\vspace{-3mm}
\begin{equation}
    \small
    \setlength{\belowdisplayskip}{0pt}
    \setlength{\abovedisplayskip}{0pt}
    S^{\Delta}_{UPF}[i][QoS] = 
    C_{UPF}[i][QoS] - S_{UPF}[i][QoS]
    \label{eq:2}
    \end{equation}

\vspace{-3mm}
\begin{equation}
    \small
    \setlength{\belowdisplayskip}{0pt}
    \setlength{\abovedisplayskip}{0pt}
    C_{UPF}[i][QoS] = \frac{ETPB_{UPF}[i]*Bytes_{UPF}[i]*\alpha[i][QoS]}{\delta}
    \label{eq:3}
    \end{equation}

\subsubsection{Compute Delay at MEC}

Similar to the abstraction considered for UPF the compute delay of a UE request at $MEC[j]$ ($D_{MEC}[j]$) is a function of MEC queue length ($Q_{MEC}[j]$), compute capacity of MEC ($C_{MEC}[j]$), and the compute server utilization ($S_{MEC}[j]$), as shown in Eqn. \ref{eq:4} and Eqn. \ref{eq:5}. Note that, unlike UPF, MEC does not distinguish among different services. The compute capacity of the MEC is modeled as the function of its CPU processing speed i.e., Execution Time Per Bit ($ETPB_{MEC}[j])$ and bytes processed in MEC per UE ($Bytes_{MEC}[j]$) shown in Eqn. \ref{eq:6}. 

\begin{equation}
    \small
    \setlength{\belowdisplayskip}{0pt}
    \setlength{\abovedisplayskip}{0pt}
        D_{MEC}[j] = (\frac{(Q_{MEC}[j] + 1 - S^{\Delta}_{MEC}[j])}{C_{MEC}[j]}+1)*\delta
    \label{eq:4}
    \end{equation}
\vspace{-1.5mm}
\begin{equation}
    \small
    \setlength{\belowdisplayskip}{0pt}
    \setlength{\abovedisplayskip}{0pt}
    S^{\Delta}_{MEC}[j] = 
    C_{MEC}[j] - S_{MEC}[j]
    \label{eq:5}
    \end{equation}
\vspace{-3mm}
\begin{equation}
    \small
    \setlength{\belowdisplayskip}{0pt}
    \setlength{\abovedisplayskip}{0pt}
    C_{MEC}[j] = \frac{ETPB_{MEC}[j]*Bytes_{MEC}[j]}{\delta}
    \label{eq:6}
    \end{equation}

\subsubsection{Network Delay}

The network delay ($D_{Net}$) of the link between an $UPF[i]$ and $MEC[j]$ can be evaluated as a function of the number of UE requests simultaneously transmitted ($N_{share}[i][j]$) between  UPF-MEC link and the corresponding link bandwidth ($BW[i][j]$), given in Eqn. \ref{eq:10}.

 \begin{equation}
    \small
    \setlength{\belowdisplayskip}{0pt}
    \setlength{\abovedisplayskip}{0pt}
     D_{Net}[i][j] =  \frac{N_{share}[i][j]*Bytes_{MEC}}{BW[i][j]*\delta}
    \label{eq:10}
    \end{equation}   

    \begin{figure}[t!]
\centering
\includegraphics[height=4.5cm, width=0.75\columnwidth]{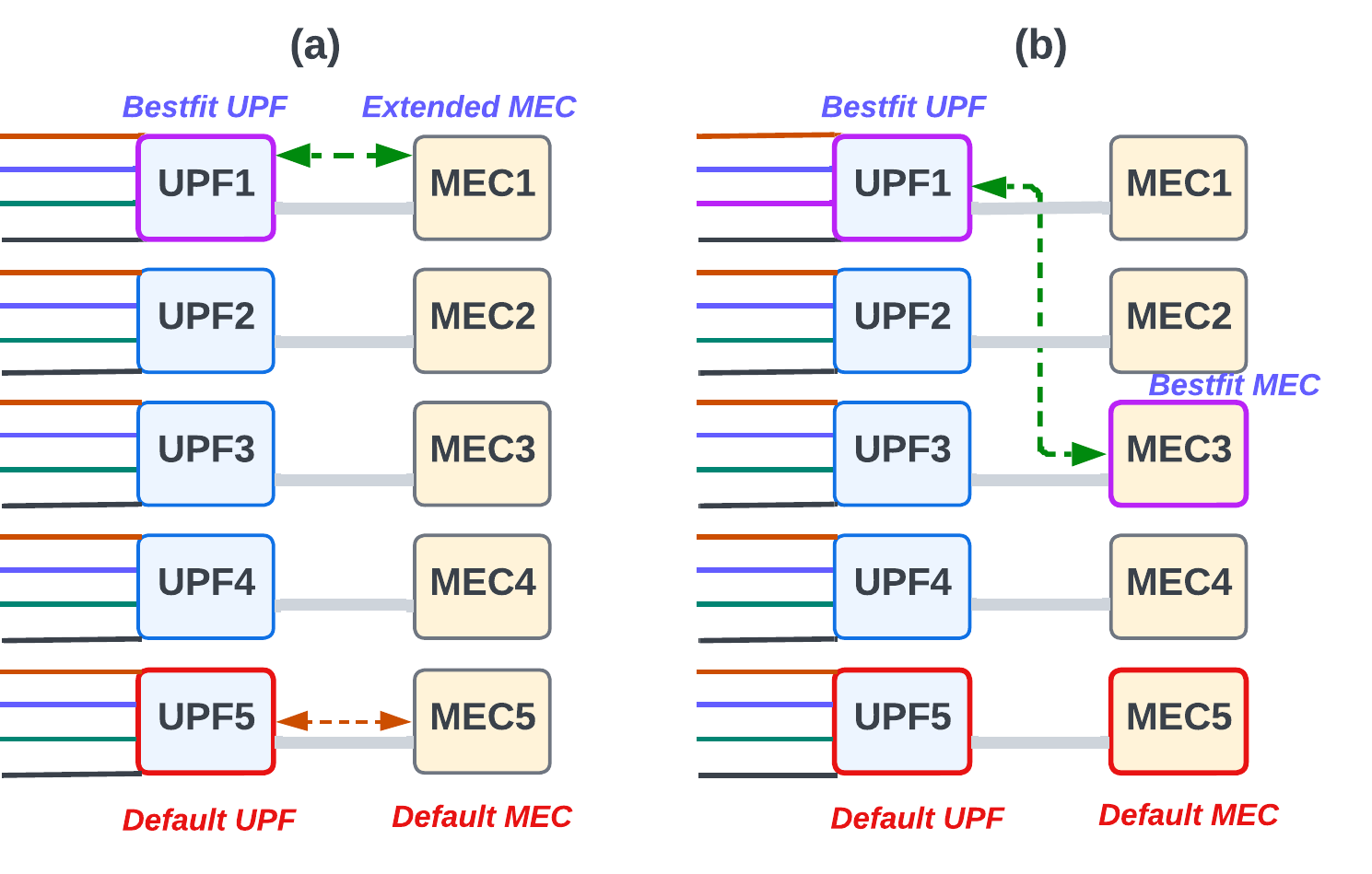}
\vspace{-2mm}
\caption{(a) Select best-fit UPF and pair with original MEC or path extension to MEC (b) Select best-fit UPF-MEC pair.}
\vspace{-3mm}
\label{fig:sim-design}
\vspace{-4mm}
\end{figure}
\vspace{-3mm}
\subsection{Problem Formulation: Obtain the bestfit UPF}
\vspace{-1mm}
Suppose, there are $N$ UE requests of specific QoS placed across the UPFs (metrics defined in Eqn. \ref{eq:1}, \ref{eq:2} and \ref{eq:3}) in the following way i.e., $x_1$ UE requests to $UPF[1]$, $x_2$ UE requests to $UPF[2]$,… $x_j$ UE requests to $UPF[j]$, … and $x_U$ UE requests to $UPF[U]$. Then considering $UPF[j]$, the worst-case compute delay (unit of epoch) is given by \\
\vspace{-3mm}
\begin{equation}
    \small
    \setlength{\belowdisplayskip}{0pt}
    \setlength{\abovedisplayskip}{0pt}
    D_{UPF}[j][QoS] = \max (0,\frac{(Q_{UPF}[i][QoS] + x_j - S^{\Delta}_{UPF}[i][QoS])}{C_{UPF}[i][QoS]})
    \label{eq:8}
    \end{equation}


Hence, the goal of the optimization problem is to find the optimal [$x_1$,..., $x_U$] to minimize the maximum of the worst-case UE compute delay of a given QoS across all UPFs. Therefore, the problem formulation is as follows: 
\vspace{-1mm}
\begin{equation}
\small
\begin{aligned}
\min_{x_1\dots x_U} \quad & \max (D_{UPF}[j][QoS])\\
\textrm{s.t.} \quad &\sum x_j = N\\
  &S_{UPF}[j][QoS] <= C_{UPF}[j][QoS]    \\
\end{aligned}
\label{eq:9}
\end{equation}

\vspace{-1mm}
\textbf{NP Hardness:} Best fit UPF allocation with one QoS type is a max-min problem. A similar formulation holds for other QoSes. The number of possible (and valid) combinations $x_j$s will increase exponentially with large $N$, making the problem computationally hard. The complexity will be of $\mathcal{O}(U^N)$ (as first UE has $U$ possibilities to place, given one placement the second UE has again $U$ possibilities to place, and so on). Hence, we propose our low-overhead heuristic where we provide per UE weighted resource allocation at UPFs, which has linear complexity w.r.t. number of UE requests.

\vspace{-2mm}
\subsection{UPF-MEC Assignment Scheme }
\label{schemes}
We evaluate the performance and effectiveness of the proposed system model under the following schemes:

\textit{1. Bestfit UPF with Path Extension to MEC (\textbf{UPF-ext)}:} This scheme extends the data path from the MEC-IA selected UPF to its corresponding MEC pair (Algorithm \ref{alg:best_fit_PE}). The \textit{Bestfit UPF}, i.e., ($BF_{UPF}$) is selected based on $min[D_{UPF}]$ from Algorithm \ref{alg:best_fit_upf}. Considering that UPF1 is selected as the $BF_{UPF}$, instead of sending the UE request from UPF 1 to default MEC 5 (selected by SMF), the path is extended to MEC 1, as seen in Figure \ref{fig:sim-design}(a) (green arrow). The path extension is an easy-to-realize design choice since the private 5G network architecture is deployed as an isolated and independent system that offers flexibility to replicate smart factory services in all the in-house MEC platforms and also set up routing and forwarding from all the UPFs to all the MECs. 

\textit{2. Bestfit UPF-MEC Pair (\textbf{UPF-MEC)}}: The MEC-IA has global view of the private 5G network, implying it is aware of the UPF and MEC topology and resource utilization.  Algorithms \ref{alg:best_fit} enables best fit UPF-MEC pair selection by invoking Algorithm \ref{alg:best_fit_upf} (best UPF) and Algorithm \ref{alg:best_fit_mec} (best MEC). As seen in Figure \ref{fig:sim-design}(b), MEC-IA extends the execution of the intelligent algorithm to the MEC layer and selects UPF 1 and MEC 3 (green arrow) as the \textit{Bestfit UPF-MEC pair}.

\begin{table}[t!]
\small
    \centering
    \begin{tabular}{|p{4.3cm}|p{4.5cm}|}
    \hline
    \textbf{Parameter} & \textbf{Value}\\
    \hline
    Number of UPFs and MECs & 5 \\
    \hline
    epoch interval & 1 ms \\
    \hline
    UPF-MEC Bandwidth & 150, 300, 450, 700, 1000 Mbps \\
    \hline
    Bytes processed at UPF, MEC  & 256 B, 1500B \\
    \hline
    UPF \& MEC processing capacity & 2.33 - 3.57 GHz \\
    \hline
    UPF QoS bucket for each service & (0.1 - 0.9)*capacity of UPF \\
    \hline
    uRLLC, eMBB latency threshold & 5, 10 ms \\
    \hline
    Compute capacity order & UPF/MEC 5 $>$ 1 $>$ 3 $>$ 2 $>$ 4 \\
    \hline
    Order of traffic arrival rate & UPF 3 $>$ 2 $>$ 5 $>$ 4 $>$ 1 \\
    \hline
    Traffic skewness & UPF 1: 13\%, UPF 2: 24\%, UPF 3: 30\%, UPF 4: 15\%, UPF 5: 18\% \\
    \hline
    \end{tabular} 
    \vspace{1mm}
     \caption{Simulation Parameters \label{tab:sim-para}} 
     \vspace{-1mm}
\end{table}

\begin{algorithm}[h]
  \caption{Bestfit UPF with Path Extension}\label{alg:best_fit_PE}
  \small
  $[BF_{UPF}, D_{UPF}[BF_{UPF}]] \gets find\_bestfit\_UPF()$

  $BF_{MEC} \gets BF_{UPF}$

      $S^{\Delta}_{MEC}[BF_{MEC}] \gets C_{MEC}[BF_{MEC}] - S_{MEC}[BF_{MEC}]$

       \If{$Q_{MEC}[BF_{MEC}] < S^{\Delta}_{MEC}[BF_{MEC}]$}
        {
         $D_{MEC}[BF_{MEC}] \gets  \delta$
         
        }
          \Else{
            \vspace{-2mm}
             $D_{MEC}[BF_{MEC}] \gets \frac{(Q_{MEC}[BF_{MEC}] + 1 - S^{\Delta}_{MEC}[BF_{MEC}])}{C_{MEC}[BF_{MEC}]}*\delta + \delta$
        }

       $D_{Net}[BF_{UPF}][BF_{MEC}] \gets \frac{N_{share}[BF_{UPF}][BF_{MEC}]*Bytes_{MEC}}{(BW[BF_{UPF}][BF_{MEC}]*\delta)}$

      $D_{E2E} \gets D_{UPF}[BF_{UPF}] + D_{Network}[BF_{UPF}][BF_{MEC}] +  D_{MEC}[BF_{MEC}]$
            
\end{algorithm}

\begin{algorithm}[h]
  \caption{Bestfit UPF-MEC pair  Selection}\label{alg:best_fit}
  \small

  $[BF_{UPF}, D_{UPF}[BF_{UPF}]] \gets find\_bestfit\_UPF()$

  $[BF_{MEC}, D_{MEC}[BF_{MEC}]] \gets find\_bestfit\_MEC()$

       $D_{Net}[BF_{UPF}][BF_{MEC}] \gets \frac{N_{share}[BF_{UPF}][BF_{MEC}]*Bytes_{MEC}}{(BW[BF_{UPF}][BF_{MEC}]*\delta)}$ 

      $D_{E2E} \gets D_{UPF}[BF_{UPF}] + D_{Net}[BF_{UPF}][BF_{MEC}] +  D_{MEC}[BF_{MEC}]$
                  
\end{algorithm}

\vspace{-3mm}

\begin{algorithm}[h]
  \caption{Function $find\_bestfit\_UPF()$}\label{alg:best_fit_upf}
\small  
      \For{$i \gets 1$ to $U$}
      {
       
      $S^{\Delta}_{UPF}[i][QoS] \gets C_{UPF}[i][QoS] - S_{UPF}[i][QoS]$

      \If{$Q_{UPF}[i][QoS] < S^{\Delta}_{UPF}[i][QoS]$}
        {
         $PC_{UPF}[i] \gets  \delta$
         
        }
        \Else{
            \vspace{-2mm}
             $PC_{UPF}[i] \gets \frac{(Q_{UPF}[i][QoS] + 1 - S^{\Delta}_{UPF}[i][QoS])}{C_{UPF}[i][QoS]}*\delta + \delta$
        }

      }
      $BF_{UPF}[i][QoS] \gets min_{index}(PC_{UPF})$
      
      $D_{UPF}[i][QoS] \gets min(PC_{UPF})$
      
   \Return [$BF_{UPF}$, $D_{UPF}[BF_{UPF}][QoS]$] 
      
\end{algorithm}

\begin{algorithm}[h]
  \caption{Function $find\_bestfit\_MEC()$}\label{alg:best_fit_mec}
    \small  
      \For{$j \gets 1$ to $M$}
      {
       
      $S^{\Delta}_{MEC}[j] \gets C_{MEC}[j] - S_{MEC}[j]$

       \If{$Q_{MEC}[j] < S^{\Delta}_{MEC}[j]$}
        {
         $PC_{MEC}[j] \gets  \delta$
         
        }
          \Else{
            \vspace{-2mm}
             $PC_{MEC}[j] \gets \frac{(Q_{MEC}[j] + 1 - S^{\Delta}_{MEC}[j])}{C_{MEC}[j]}*\delta + \delta$
        }

      }
      
      $BF_{MEC} \gets min_{index}(PC_{MEC})$

      $D_{MEC}[BF_{MEC}] \gets min(PC_{MEC})$
      
   \Return [$BF_{MEC}$, $D_{MEC}[BF_{MEC}]$] 
      
\end{algorithm}

\vspace{-4mm}
\section{Results and Discussion}

The simulation parameters are listed in Table \ref{tab:sim-para}. A flow-level simulator was developed that simulates various resource management schemes in UPF and MEC layers 
(Section \ref{schemes}).
\vspace{-3mm}

\begin{figure*}[t]
\centering
\includegraphics[width=\textwidth]{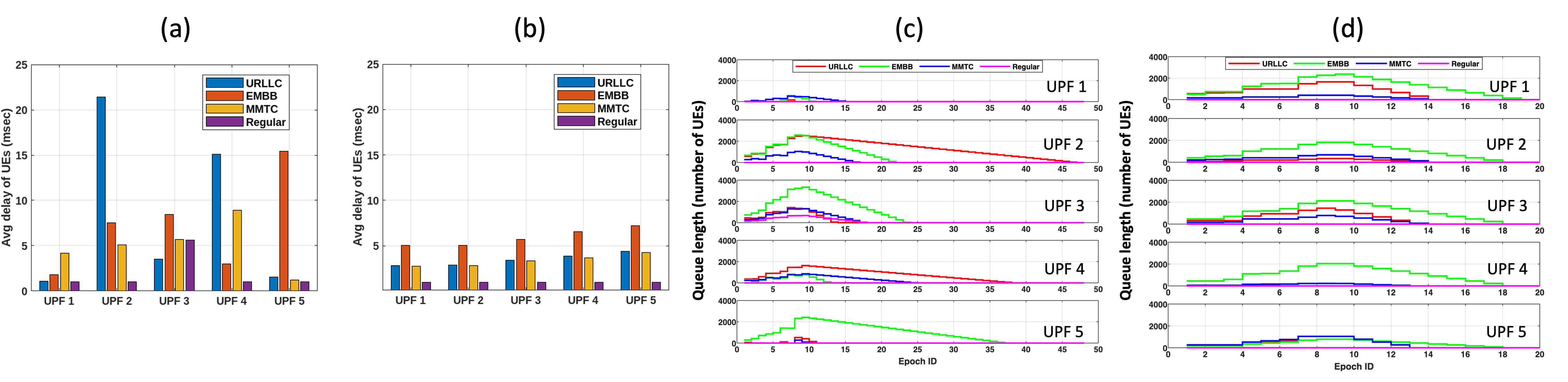}
\vspace{-7mm}
\caption{Average data plane delay for different slices across all UPFs for (a) baseline, (b) MEC-IA and corresponding  queue length (c) baseline, (d) MEC-IA }
\label{fig:UPF}
\vspace{-2mm}
\end{figure*}
\vspace{-1mm}
\subsection{Data Plane Delay in UPF across Diverse Service Types}

Figure \ref{fig:UPF}(a) depicts performance of uRLLC, eMBB, mMTC, and regular traffic at each UPF in a smart factory for the baseline scenario. Notably, uRLLC requests at UPF 2 face a high average delay of 22 ms due to SMF's UE assignment to the UPF based on location and load, but not computing availability per QoS type. In contrast, Figure \ref{fig:UPF}(b) shows MEC-IA's dynamic resource management, optimizing compute capacity for each QoS traffic type across all UPFs. For instance, UPF 2 and UPF 4 struggle with uRLLC requests, and UPF 5 with eMBB requests due to the demanding nature of theses services, necessitating dedicated resources for sub-10 ms latency. MEC-IA leverages under-utilized resources, proactively assigning incoming UE requests, thereby reducing average latency at UPFs for all traffic types. In Figure \ref{fig:UPF}(b), all the services across the UPFs record an average latency of $<$ 10 ms. Analyzing the queue lengths in Figure \ref{fig:UPF}(c), we see high peaks at UPF 2 and UPF 3 due to traffic skewness (refer Table \ref{tab:sim-para}). The spike in average delay for uRLLC requests in UPF 2 is captured by the time taken by uRLLC traffic to drain from the queue of UPF 2, $\approx$ 46 ms. Similar is the case with other UPFs in the baseline. While the MEC-IA enabled resource distribution allows close to similar queue peak and drain time (Figure \ref{fig:UPF}(d)).



\subsection{Extending Data Plane Resource Management to MEC }
\vspace{-1mm}

\begin{figure}[t]
\centering
\includegraphics[width=\columnwidth]{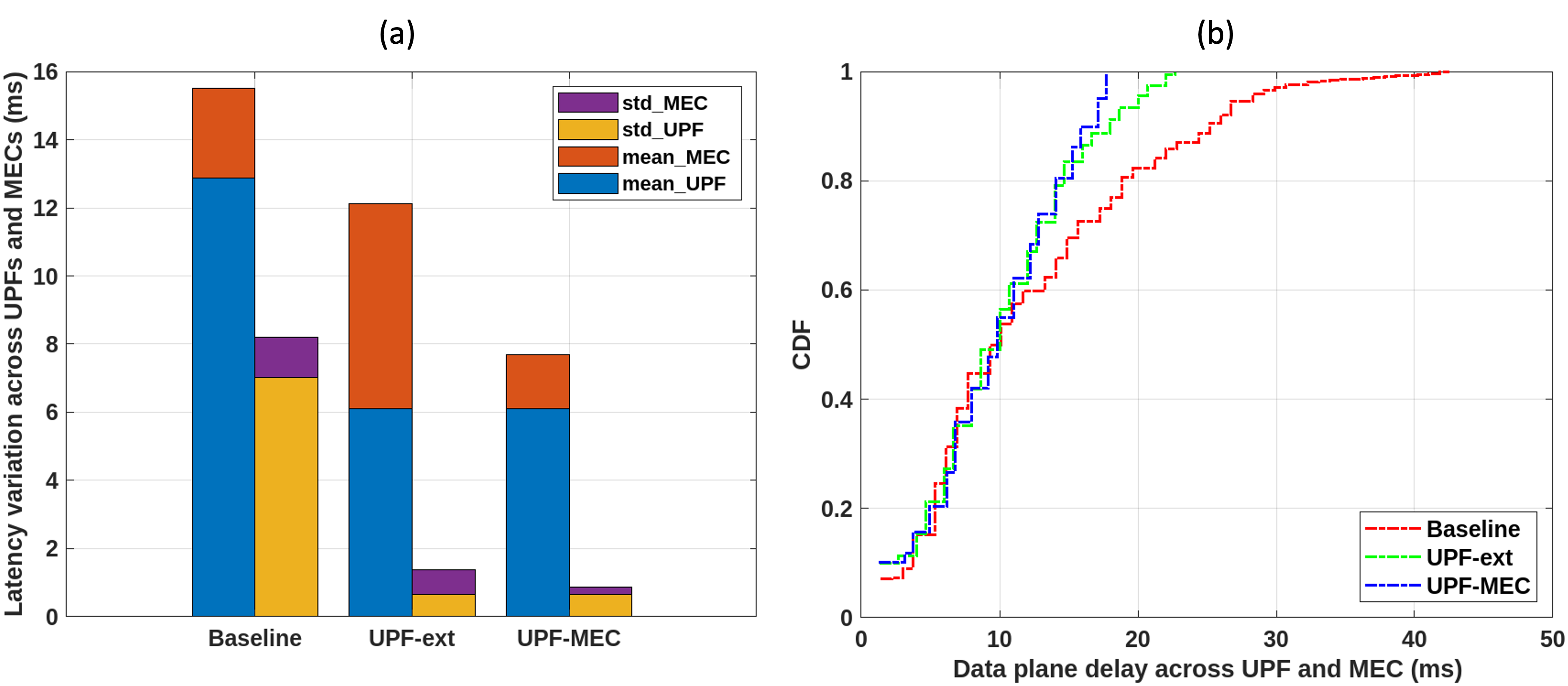}
\vspace{-6mm}
 \caption{(a) Average and standard deviation of delay across UPFs and MECs for all the schemes (b) Combined data plane delay at UPF and MEC.}
\label{MEC-Combined}
\vspace{-5mm}
\end{figure}

Figure \ref{MEC-Combined}(a) demonstrates the efficacy of extending  MEC-IA functionality to the MEC layer while processing UE requests. Previous approaches consider furnishing MEC computational resources in close proximity to the end-user \cite{UE-MEC, UE-MEC1}, enabling low latency. However, such benefit could be nullified by the resource bottleneck across the MEC platforms (Figure \ref{MEC-Combined}(a)). 
In comparison to the baseline, MEC-IA assisted UPF-ext scheme reduces both the average UPF latency (by $2\times$, blue bar stack) and standard deviation (by $10.7\times$, yellow bar stack). However, the average latency across MECs surges by $2.35\times$ in the UPF-ext scheme (red bar stack), reducing the overall benefit. Hence, it is imperative to extend the MEC-IA assisted dynamic resource management to the MEC layer (UPF-MEC scheme). The average MEC latency and STD are reduced by $1.63\times$ and $5.2\times$ (violet bar stack) w.r.t. baseline, indicating the optimal matching of a UE request to a bestfit MEC.

\vspace{-1mm}
\subsection{Evaluation of Data Plane Delay across UPF-MEC}
\vspace{-1mm}
Figure \ref{MEC-Combined}(b) represents the cumulative 
probability distribution of the delay experienced by the end users in the proposed system. In this context, UPF-MEC data plane delay implies the delay experienced by user requests across the UPFs and MEC platforms due to compute and network resource bottlenecks. The baseline curve grows gently and records a worst-case latency of $\approx$ $42$ ms,  while $80$ percentile latency is $<$ 20 ms. The Bestfit UPF with no path extension scheme records a worst-case UPF-MEC latency of $27$ ms, marking a $36\%$  delay reduction compared to the baseline approach. 
Next, considering the bestfit UPF with path extension scheme, the worst-case latency reduces by $48\%$ and $18.5\%$ in comparison to baseline and No-PE schemes respectively. Finally, for the best-fit UPF-MEC scheme, the CDF curve shows a steeper rise compared to other schemes indicating a larger proportion of UE connections experiencing lower delays. We can observe that the tail latency reduces by $57\%$ compared to the baseline, $33\%$ compared to no path extension scheme and $21.7\%$ compared to path extension scheme. Also, the $80$ percentile latency recorded is less than 15 ms. 


\vspace{-5mm}
\begin{figure}[h]
\centering
\includegraphics[width=\columnwidth]{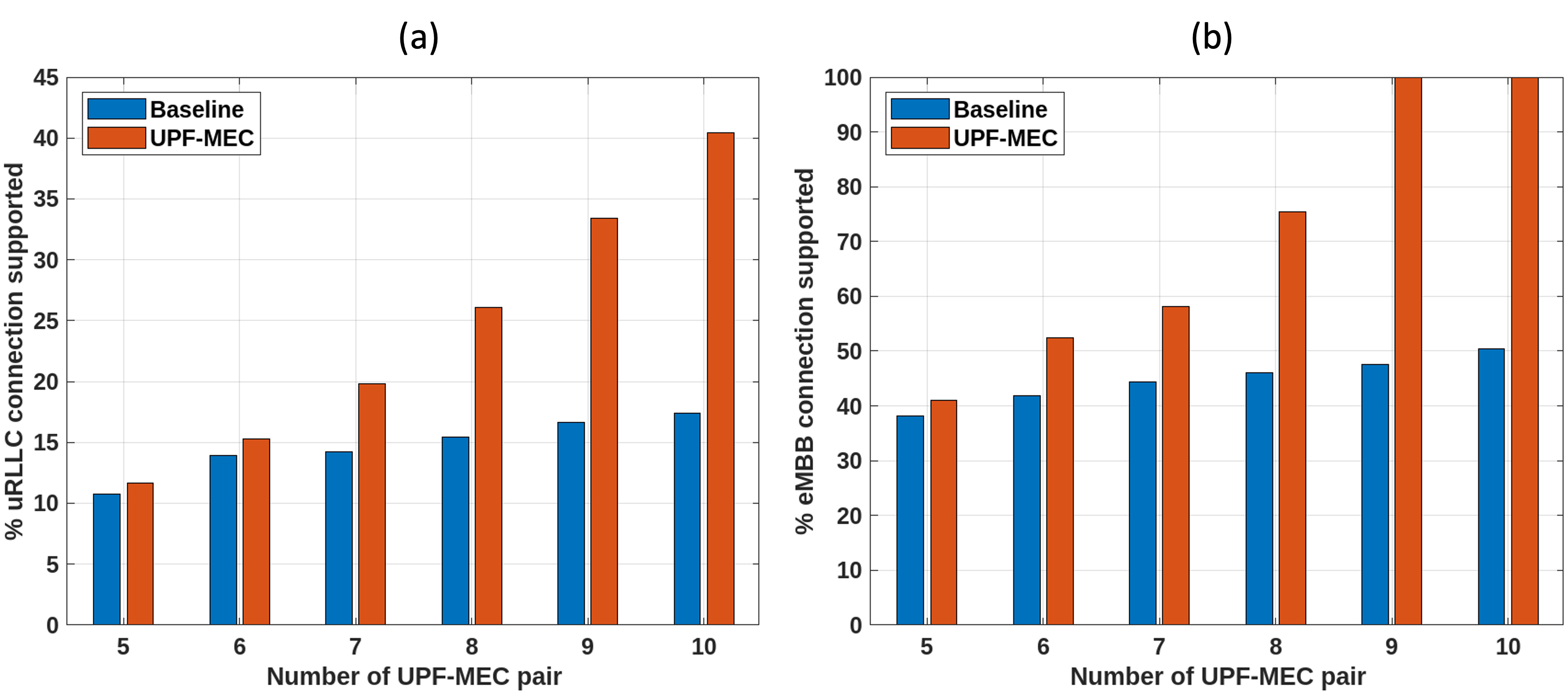}
\vspace{-5mm}
\caption{(a) Percentage of uRLLC connections with UPF and MEC data plane delay below $5$ ms for different schemes. (b) Percentage of eMBB connections with UPF and MEC data plane delay below $10$ ms for different schemes.}
\vspace{-3mm}
\label{fig:capex_urllc}
\end{figure}

\vspace{-2mm}
\subsection{Resource CapEx Analysis}

Figures \ref{fig:capex_urllc}(a) and (b) demonstrate that our MEC-IA assisted proactive resource management can effectively improve the QoE for diverse applications while saving network resources. Another interpretation of such benefit is that our scheme can potentially reduce the resource CapEx for private 5G deployed in future factory ecosystems. 
As shown in Figure \ref{fig:capex_urllc} (a), we increase the number of UPF-MEC pair and observe the percentage of uRLLC connections meeting the end-to-end data plane latency below the designated threshold ($5$ msec). For the baseline, $17\%$ of the uRLLC connections are able to meet the $5$ ms threshold with $10$ UPF-MEC pairs, whereas almost similar percentage ($15\%$) can be met with $6$ UPF-MEC pairs in MEC-IA scheme. This effectively implies $40\%$ CapEx savings, i.e., for every $10$ units (each UPF-MEC pair costs $1$ unit) spent in baseline,  the MEC-IA would achieve the same performance (\% URLLC connections) by spending only $6$ units. Further, it can be observed that for a CapEx investment of $10$ units (10 UPF-MEC pairs), the baseline can support $17\%$ of uRLLC connections, whereas MEC-IA can support $41\%$ of uRLLC connections under the $5$ ms threshold, recording a uRLLC connectivity gain of $\approx$ $2.40\times$. 
A similar trend can be observed for eMBB connections  (Figure \ref{fig:capex_urllc}(b)). For a CapEx investment
of $9$ units ($9$ pairs of UPF-MEC), the baseline system can support $48\%$ of eMBB connections. On the contrary, MEC-IA can support $100\%$ of eMBB connections under the $10$ ms threshold, recording an eMBB connectivity gain of $\approx$ $2.08 \times$.

\vspace{-2mm}
\section{Related Work}
The work in \cite{hyperv} presented a preliminary
design for HyperVDP, a P4-specific programmable data
plane; \cite{hyperv2} implements the dynamic controller supporting instantiation of NFs. Similar research e.g. \cite{FPGA}, \cite{FPGA2}, provide both high performance and programmability based on the FPGA programmable data plane. Compute-capable I/O devices, such as smart NICs/SSDs \cite{smartnic2, smartnic}, and accelerators \cite{acc, acc2} are used to offload data plane operations. 
Our proposed scheme is precursory to offloading works, as it already optimizes the UE to UPF/MEC placements before data plane processing commences. Combining such resource-aware proactive UE placement with dataplane offloading would be interesting future work.

In \cite{SDN}, an SDN controller orchestrates load balancing to ensure network flows for low-latency communications. In \cite{HULA}, a centralized load balancing technique achieves equal-cost multi-path routing in the data plane. However, this scheme is congestion agnostic and only splits traffic at the flow level, exhibiting degraded performance during link failures. On the contrary, our MEC-IA is a proactive solution that assists the control plane network functions in selecting the best-fit data plane component, the UPF, and setting up the PDU session to cater to UE traffic of varied QoS types.

\vspace{-1mm}
\section{Conclusion and future work}
\vspace{-1mm}


This paper introduces MEC-IA, an intelligent real-time resource management framework for cloud-native private 5G network core, to alleviate the processing bottleneck at the dataplane. MEC-IA, hosted in mobile edge computation platform, enables dynamic and proactive resource allocation that can enhance QoS for performance-critical applications while significantly improving system utilization. In our future work, we aim to integrate MEC-IA with the application programmable interfaces (APIs) into our private 5G deployment within the COSMOS environment. This extension will involve utilizing the most recent open-source 5G core (introduced in May, 2023) that features a 3GPP standardized NEF component.


\vspace{12pt}


\vspace{-2mm}
\begin{thebibliography}{00}

\bibitem{COSMOS}Raychaudhuri, Dipankar, et al. "Challenge: COSMOS: A city-scale programmable testbed for experimentation with advanced wireless." Proceedings of the 26th Annual International Conference on Mobile Computing and Networking. 2020.
\bibitem{Accel}A. Bose et al., (2022, October). AccelUPF: accelerating the 5G user plane using programmable hardware. In Proceedings of the Symposium on SDN Research (pp. 1-15).

\bibitem{SDN-NFV}Li, Y., \& Chen, M. (2015). Software-defined network function virtualization: A survey. IEEE Access, 3, 2542-2553.

\bibitem{science-direct}Plane function. Plane Function - an overview | ScienceDirect Topics. https://www.sciencedirect.com/topics/computer-science/plane-function (Access Date: Sep 2023)

\bibitem{HULA}Katta, N., Hira, M., Kim, C., Sivaraman, A., \& Rexford, J. (2016, March). Hula: Scalable load balancing using programmable data planes. In Proceedings of the Symposium on SDN Research (pp. 1-12).

\bibitem{HULA2}Hsu, K. F., et al., (2020, March). Adaptive weighted traffic splitting in programmable data planes. In Proceedings of the Symposium on SDN Research (pp. 103-109).

\bibitem{usecase1}M. Chen et al., “Data-driven computing and caching in 5G networks: Architecture and delay analysis,” IEEE Wireless Commun., vol. 25, no. 1, pp. 70–75, Feb. 2018.

\bibitem{amarisoft}{“Software Company Dedicated to 4G LTE and 5G Nr.” Amarisoft, 8 June. 2023, https://www.amarisoft.com/.}


\bibitem{N3-N9}TSG SA, System Architecture for the 5G System (5GS); Stage 2,
document TS 23.501 V16.4.0, 3GPP, Mar. 2020.

\bibitem{UE-MEC}Emara, M., Filippou, M. C., \& Sabella, D. (2018, June). MEC-assisted end-to-end latency evaluations for C-V2X communications. In 2018 European conference on networks and communications (EuCNC) (pp. 1-9). IEEE.

\bibitem{UE-MEC1}Makris, N., Passas, V., Nanis, C., \& Korakis, T. (2019, July). On minimizing service access latency: Employing mec on the fronthaul of heterogeneous 5g architectures. In 2019 IEEE international symposium on local and metropolitan area networks (LANMAN) (pp. 1-6). IEEE.

\bibitem{hyperv}C. Zhang et al., “HyperV: A high performance hypervisor for virtualization of the programmable data
plane,” in Proc. 26th Int. Conf. Comput. Commun. Netw. (ICCCN), Jul. 2017, pp. 1–9.

\bibitem{hyperv2} C. Zhang et al., Hyper Project. Accessed: May 2018. [Online]. Available: https://github.com/HyperVDP.

\bibitem{FPGA}B. Li et al., “ClickNP: Highly flexible and high-performance network processing with reconfigurable hardware,” in Proc. ACM SIGCOMM Conf., New York, NY, USA, 2016, pp. 1–14, doi:
10.1145/2934872.2934897.

\bibitem{FPGA2}Y. Liao, D. Yin, and L. Gao, “PDP: Parallelizing data plane in virtual network substrate,” in Proc. 1st ACM Workshop Virtualized Infrastruct. Syst. Archit. (VISA), New York, NY, USA, 2009, pp. 9–18, doi: 10.1145/1592648.1592651.


\bibitem{smartnic2}D. Firestone et al., “Azure accelerated networking: Smartnics in the public cloud,” in 15th USENIX Symposium on Networked Systems Design and Implementation (NSDI 18).
Renton, WA: USENIX Association, Apr. 2018, pp. 51–66.

\bibitem{smartnic}M. Liu, T. Cui, H. Schuh, A. Krishnamurthy, S. Peter, and K. Gupta, “Offloading distributed applications onto smartnics using ipipe,” in Proceedings of the ACM Special Interest Group on Data Communication, ser. SIGCOMM ’19. New York, NY, USA: ACM, 2019, pp. 318–333.

\bibitem{acc} Y. Go, M. Jamshed, Y. Moon, C. Hwang, and K. Park, “Apunet:
Revitalizing gpu as packet processing accelerator,” in Proceedings
of the 14th USENIX Conference on Networked Systems Design and
Implementation, ser. NSDI’17. USA: USENIX Association, 2017, p.
83–96

\bibitem{acc2} S. Han, K. Jang, K. Park, and S. Moon, “Packetshader: A GPU accelerated software router,” in Proceedings of the ACM SIGCOMM
2010 Conference, ser. SIGCOMM ’10. New York, NY, USA:
Association for Computing Machinery, 2010, p. 195–206.

\bibitem{SDN} Filali, Abderrahime, et al. "Preemptive SDN load balancing with machine learning for delay sensitive applications." IEEE Transactions on Vehicular Technology 69.12 (2020): 15947-15963.





\end{thebibliography}
\end{document}